\documentclass[
    ,final            
  ]
  {aipproc}

\layoutstyle{6x9}

\begin{document}

\title{Measurement of the production cross section for W-- 
and Z--bosons in association with jets in ATLAS}


\author{Stefan Ask \\ (on behalf of the ATLAS collaboration)}{
  address={Cavendish Laboratory, University of Cambridge, Cambridge, UK.}
}

\begin{abstract}
We report on the measurements of inclusive $W$+jets and $Z$+jets 
cross sections in proton--proton collisions at $\sqrt{s} = 7$ 
TeV with the ATLAS detector. Cross sections, in both the electron 
and muon decay modes of the bosons, are presented as a function 
of jet multiplicity, the transverse momentum of the jets and the 
quantity $H_T$ which is the scalar sum of the $p_T$ in the event. 
Measurements are also presented of the ratios of cross sections. 
The measured cross sections are compared to different particle--level 
predictions, based on perturbative QCD, where the measured $W$+3jet 
cross section is for the first time compared with next--to--leading 
order calculations. 
\end{abstract}

\maketitle


\section{Introduction}

The experimentally clean signatures of $W$-- and $Z$--bosons 
make the measurement of these processes in association 
with jets well suited to test perturbative QCD at the LHC. 
The processes allow for comparisons of multi--jet production 
with predictions either from the parton shower approach or 
from exact multi--parton matrix elements ($ME$) matched with 
parton showers. In addition, full next--to--leading order 
($NLO$) calculations are also available for comparison with 
many of the results. The $W/Z$ processes also differ from 
pure QCD multi--jet processes with respect to the scale of 
the hard interaction, due to the large mass of the electroweak 
gauge bosons.
 
Measurements of $W$/$Z$+jets are also important to control 
backgrounds to other measurements at the LHC. In the Standard 
Model context, one example is the top quark cross section 
measurements, where $W$+jet is often the dominant background. 
Also several beyond the standard model searches, such as the 
zero lepton SUSY search, suffer from irreducible background 
from either $W$+jets or $Z$+jets, or both.

Here we report on the ATLAS $Z$+jets and $W$+jets cross section 
measurements \cite{bib:confZ,bib:confW} based on data recorded 
during 2010. The analyses include the electron and muon 
decay channels and are based on an integrated luminosity of 
33 pb$^{-1}$.

\section{ATLAS Analysis}

The ATLAS detector systems were all fully operational during 
this data taking period and the detector acceptance considered 
was approximately determined by the following constraints. 
Electrons were used within the inner detector acceptance 
($|\eta| < 2.5$), whereas reconstructed muons were considered 
inside the acceptance of the trigger chambers ($|\eta|<2.4$). 
The electron (muon) $E_T$ ($p_T$) was required to be larger 
than 20 GeV, in order to be well inside the highly efficient 
plateau of the triggers. Jets were reconstructed inside the 
main calorimeters ($|\eta| < 3.2$) and missing transverse energy 
was based on the full calorimeter acceptance ($|\eta|<4.5$). 
The cross section measurements were made within the kinematic 
region defined by the event selection, which was well covered 
by the ATLAS detector acceptance,
\begin{itemize}
\item
 $Electrons: ~E_T > 20 ~\mbox{GeV}; ~|\eta|<2.47; ~excluding ~1.37<|\eta|<1.52.$
\item
 $Muons: ~p_T > 20 ~\mbox{GeV}; ~|\eta|<2.4.$
\item
 $Jets ~(anti-k_T, ~R=0.4): ~p_T > 20 ~or~ 30 ~\mbox{GeV}; ~|y ~or~ \eta|<2.8; ~\Delta R(\ell, jet) > 0.5.$
\item
 $W ~selection: N_{\ell} = 1; ~E_T^{miss}>25 ~\mbox{GeV}; ~m_T > 40 ~\mbox{GeV}.$
\item
 $Z ~selection: N_{\ell} = 2; ~66 < m_{\ell \ell} < 116 ~\mbox{GeV}.$
\end{itemize}
Note that here the jet $p_T$ requirement, as well as the rapidity 
variable, differs between the $W$ (20 GeV) and $Z$ analysis (30 GeV). 
The $\Delta R(\ell, jet)$\footnote{The following definition was 
used, $\Delta R = \sqrt{\Delta \eta ^2 + \Delta \phi ^2}.$} criteria 
refers to the leptons and all selected jets and the $Z$ selection 
also require the two leptons to be of opposite charge. The results 
were then corrected for detector effects, using a bin--by--bin 
unfolding method, and compared with theory ($MC$) predictions 
inside the same kinematic region. For the theory results, jets 
were reconstructed using the same algorithm based on all final 
state particles with a lifetime larger than 10 ps, except the 
leptons from the $W/Z$ decays. Lepton momenta also included any 
photons radiated within $\Delta R < 0.1$.

A good agreement was generally found between the selected data 
candidates and predictions from MC, for both $Z$ and $W$ events 
in both the electron and muon channels. Regarding the background for 
these measurements, the background coming from QCD was estimated 
based on a data--driven method whereas the electroweak and top 
backgrounds were estimated from MC. The background contamination 
of the selected $Z$+jets samples was of the order of 1\% for the 
muon channel, as well as 5\% for the electron channel. In the 
$W$+jets samples, the background was in the order of 10\%. The 
main source of uncertainty in these measurements comes from the 
jet energy scale, which contributes with approximately 10\%, 
followed by pile--up corrections ($\sim 5$\%) and luminosity 
($\sim 4$\%). More details about the analysis are found in 
\cite{bib:confZ,bib:confW}.

\section{Cross Section Measurements}

The obtained number of events were then used to measure the 
differential cross section times branching ratio with respect 
to a number of different quantities. All the results correspond 
to inclusive measurements, corrected for detector effects, 
within the kinematic region defined by the event selection 
above. 

\begin{figure}[h]
  \includegraphics[height=.47\textheight]{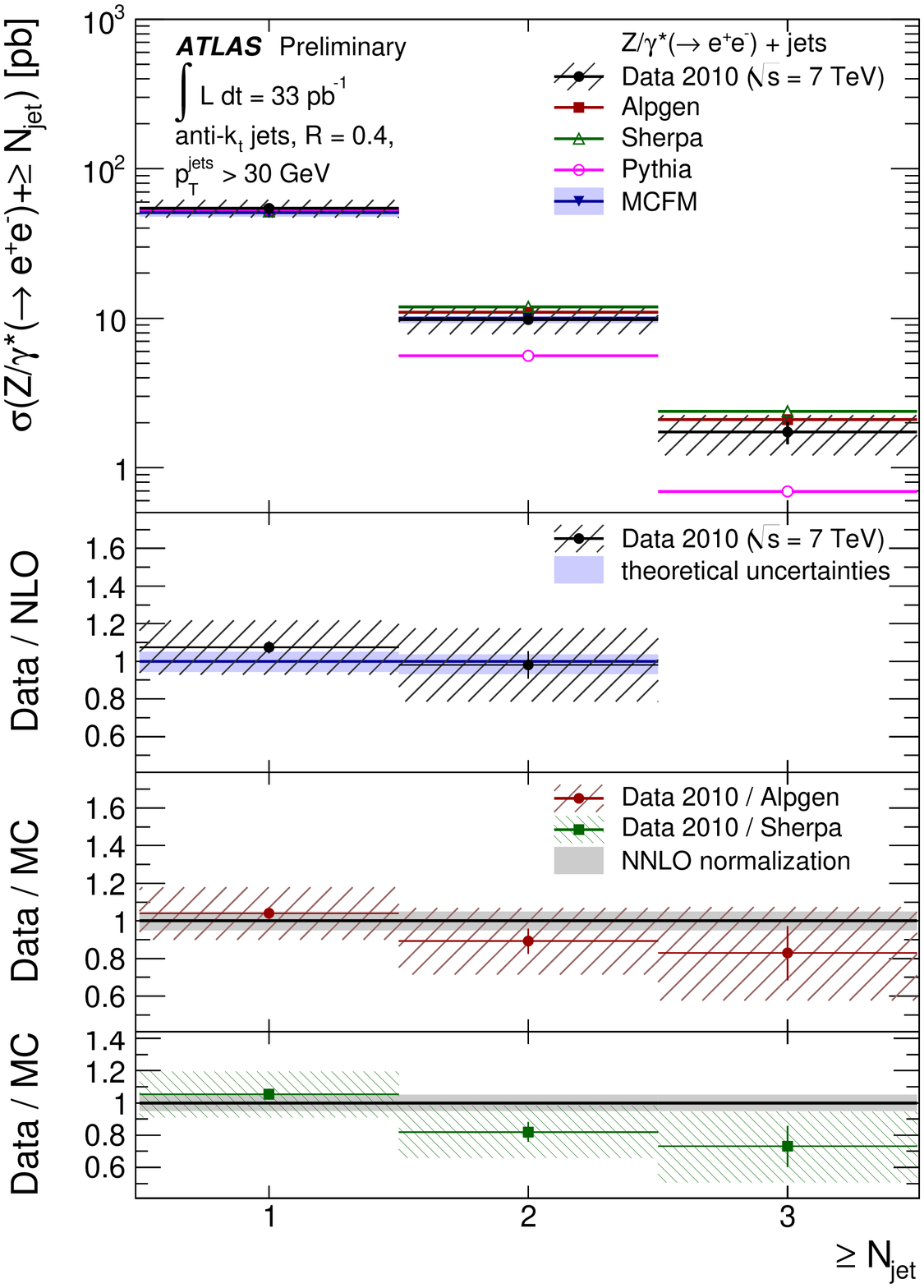}
  \includegraphics[height=.47\textheight]{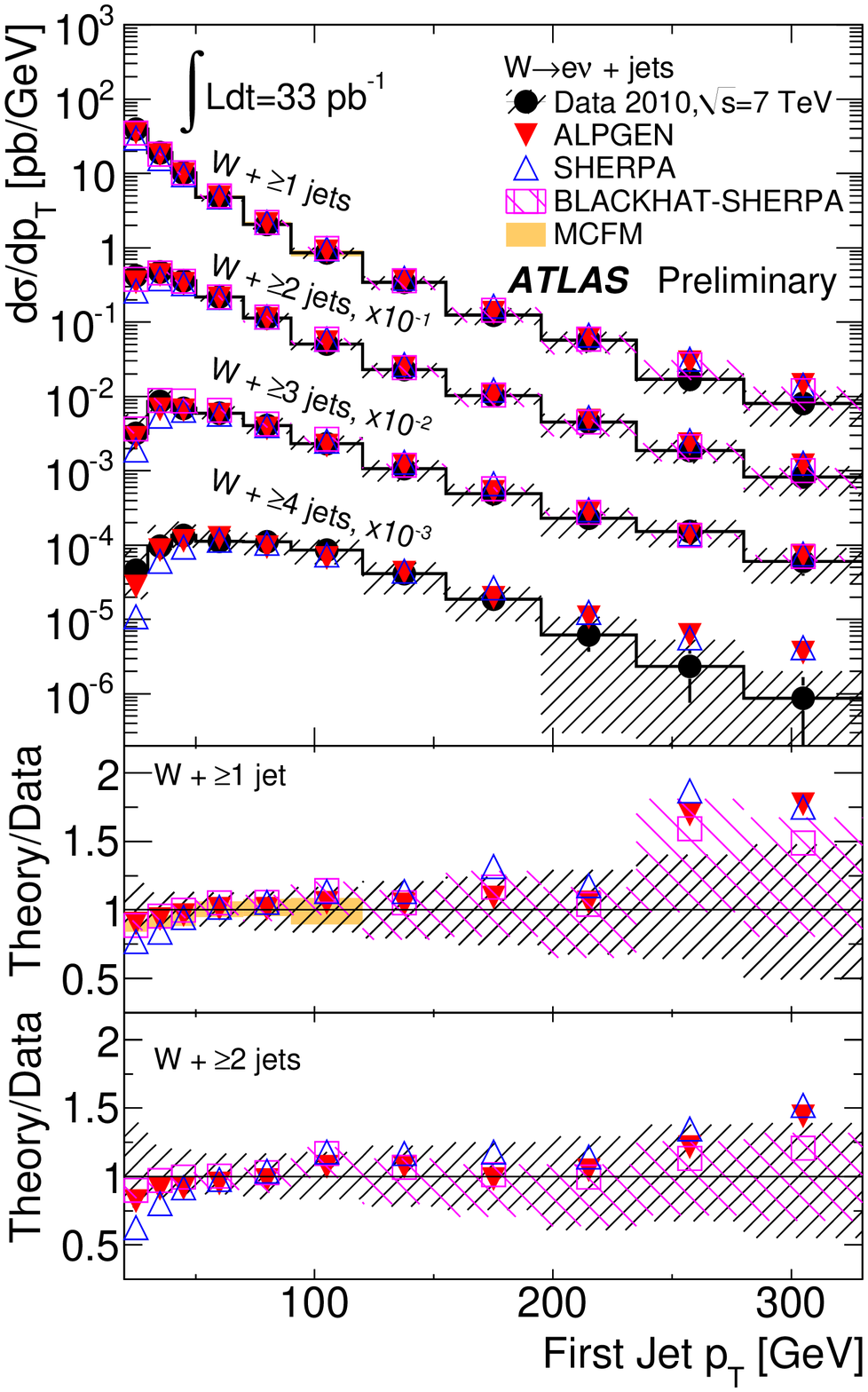}  
  \caption{Cross section for $Z \rightarrow e e$ as function of 
    the number of jets (left). Cross section for $W \rightarrow e \nu$ 
    as function of $p_T$ of the leading jet (right).\label{fig:zw2}}
\end{figure}

The differential cross section with respect of the number of 
selected jets was measured both using $W$ and $Z$ events. The 
absolute cross section and the ratio of cross sections from 
events with $N$ jets over $N - 1$ jets were measured. Some of 
the uncertainties are reduced in the ratios. Figure~\ref{fig:zw2} 
(left) shows the $Z \rightarrow e e$ cross section as a function 
of number of selected jets. The measured values show a good 
agreement with the NLO predictions, here represented by results 
obtained by {\sc MCFM}. The results are also in good agreement 
with expectations from the multi--parton ME programs ({\sc Alpgen} 
and {\sc Sherpa}), which have been normalized to the inclusive 
NNLO cross sections obtained by the FEWZ program. The results 
do on the other hand show poor agreement with the LO plus parton 
shower results ({\sc Pythia}) for events with more than one jet. 
This is due to the combination of a not--properly--covered phase 
space ($Z$s are not produced by the parton shower) together with 
using the parton shower approach for the hard jets. The results 
are shown together with the corresponding ratios between the 
results obtained from data over predictions from the MC programs 
{\sc MCFM}, {\sc Alpgen} and {\sc Sherpa}.

\begin{figure}[h]
  \includegraphics[height=.47\textheight]{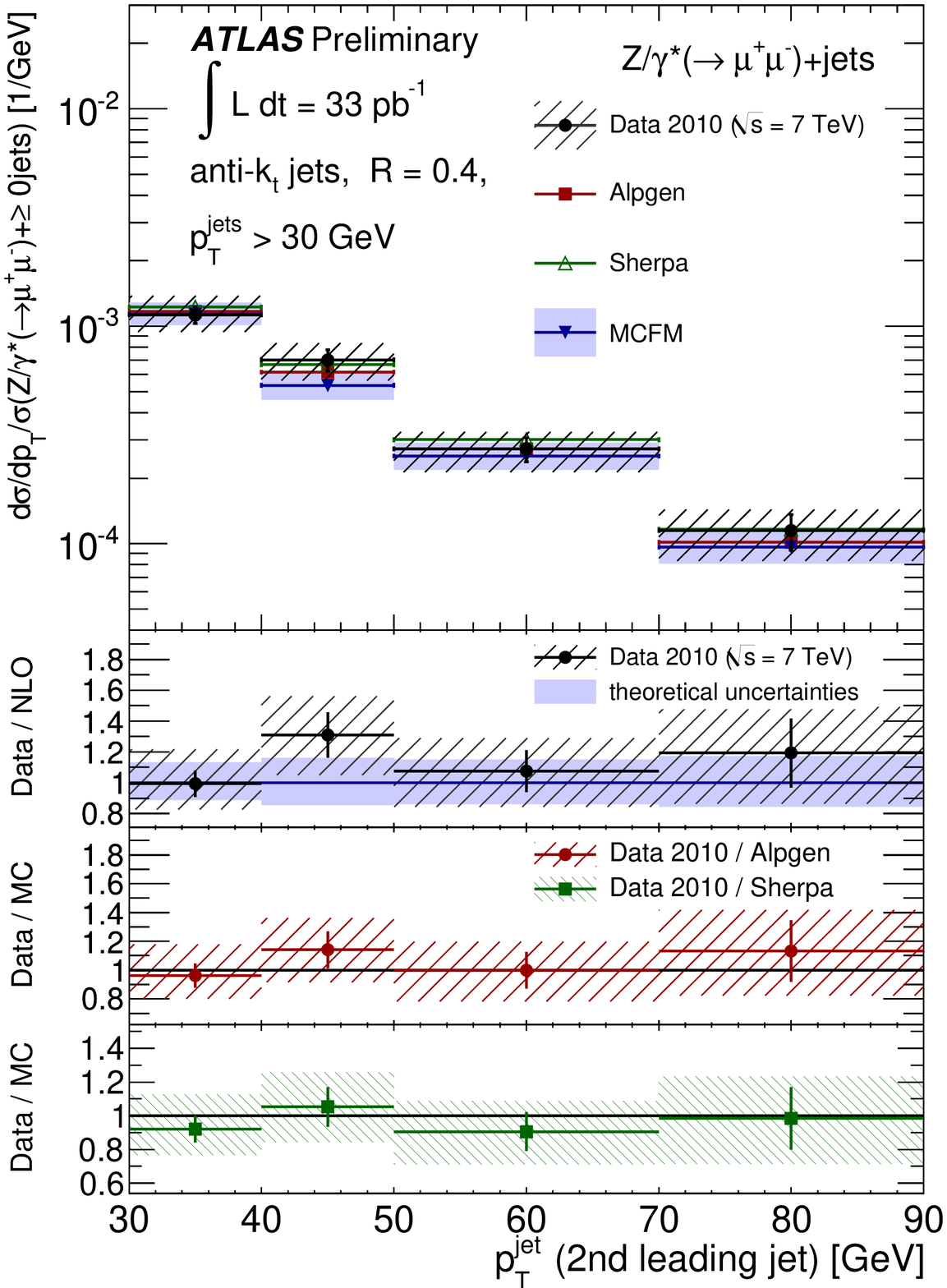}
  \includegraphics[height=.47\textheight]{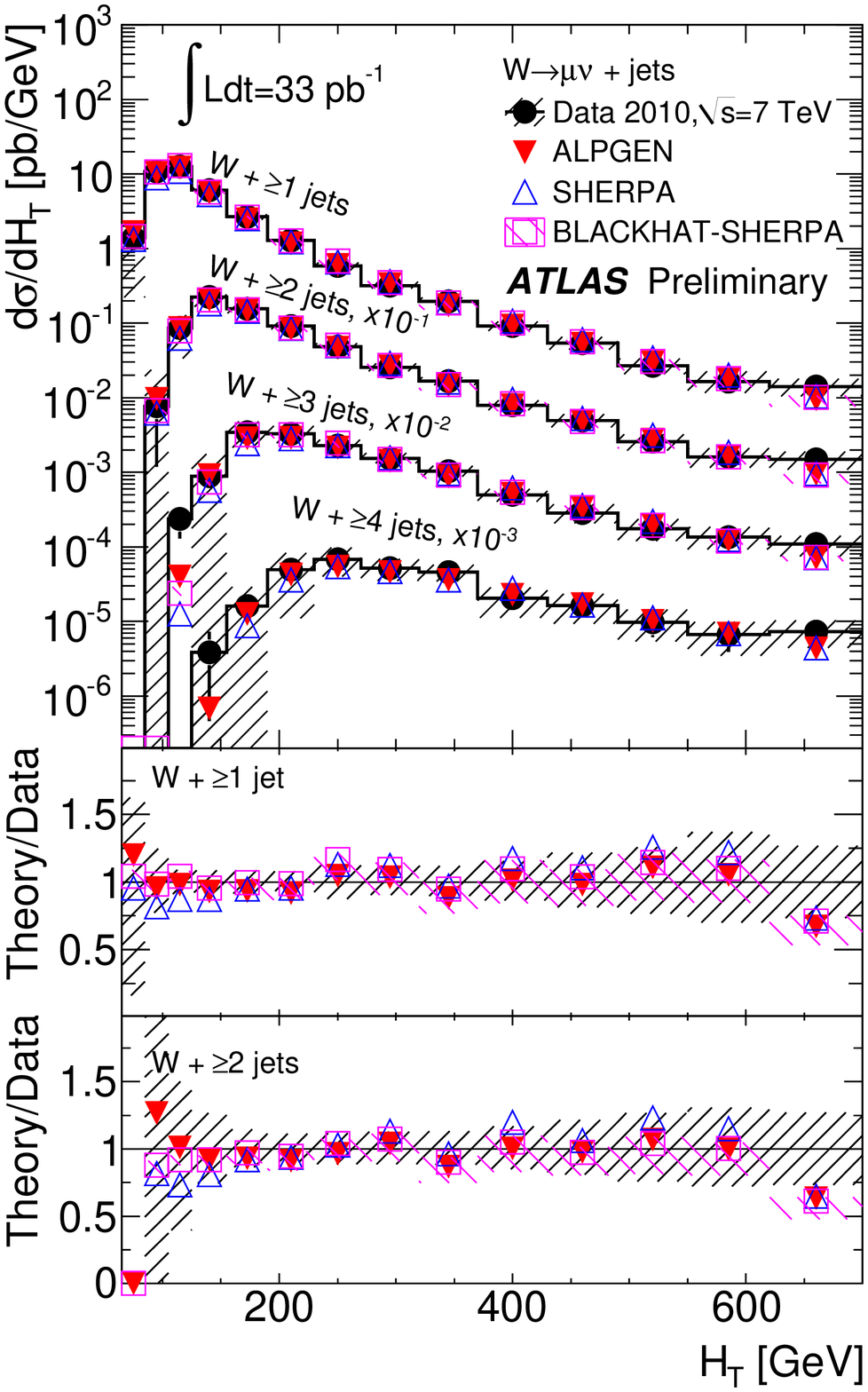}  
  \caption{Cross section for $Z \rightarrow \mu \mu$ as function 
    of $p_T$ of the second leading jet (left). Cross section for 
    $W \rightarrow \mu \nu$ as function of $H_T$ (right).\label{fig:zw3}}
\end{figure}

Figure~\ref{fig:zw2} (right) shows the differential cross section 
with respect to the leading jet $p_T$ for $W \rightarrow e \nu$. 
The measurement is performed separately for events with 1 to 4 
jets. The results from the $W$+jets measurements are also compared 
against NLO predictions from {\sc Blackhat--Sherpa}, where $W$+3jets 
predictions at NLO is compared with LHC data for the first time. 
The results are shown together with the corresponding ratios between 
results from MC over data, for events with 1 and 2 selected jets. 
Again a good agreement was found between the measurement and the MC 
predictions. The differential cross sections were also measured with 
respect to the $p_T$ of the other selected jets in the event, using 
the 2 leading jets in $Z$ events and up to 4 leading jets in the $W$ 
analysis. The cross section with respect to the $p_T$ of the second 
leading jet is shown in figure~\ref{fig:zw3} (left) for $Z \rightarrow \mu \mu$. 
A good agreement is shown and similar agreements were also found with 
respect to the other jets, both in the $W$ and $Z$ results.

In the $W$ analysis the cross section was also measured with respect 
to $H_T$. This quantity corresponds to the scalar sum of the $p_T$ 
from the jets as well as leptons, i.e. muon, electron and neutrinos, 
in the event, and this is the quantity which was used to characterize 
the scale of the hard process in the MC simulations. The results from 
$W \rightarrow \mu \nu$ are shown in figure~\ref{fig:zw3} (right), 
which are produced separately for event with 1 to 4 jets. The ratios 
between the results from MC over data are shown and a generally good 
agreement was found in both the electron as well as the muon channel.

\bibliographystyle{aipproc}   

\end{document}